# The Landau-Migdal parameter and quenching factor from the analysis of the Gamow-Teller strength distributions in medium-heavy-mass doubly-closed-shell parent nuclei

V.I. Bondarenko[1], M.H. Urin[2]

.[1]A.V. Shubnikov Institute of Crystallography of the Kurchatov Complex Crystallography and Photonics of the NRC "Kurchatov Institute", Moscow, Russia

[2]National Research Nuclear University MEPhI (Moscow Engineering Physics Institute), Moscow, Russia

The Landau-Migdal parameter g' and quenching factor Q for the Gamow-Teller strength distributions in the $^{48}$Ca, $^{90}$Zr, $^{132}$Sn and $^{208}$Pb parent nuclei are deduced from a detailed comparison of experimental distributions (obtained for the $\beta^{(-)}$ -channel) with respective strength functions evaluated within the semi-microscopic Particle-Hole Dispersive Optical Model. The deduced values are compared with the values obtained in other studies.

In the presented study, we consider a possibility to deduce from proper experimental data the values of two phenomenological quantities related to nuclear physics and its applications to astrophysics: the (dimensionless) strength g' of the spin-isospin part of Landau-Migdal p-h interaction and the quenching factor Q, which describes a suppression of the Gamow-Teller (GT) strength due to coupling to non-nucleonic degrees of freedom. Both quantities determine the GT strength distributions described within microscopic (or semi-microscopic) nuclear models, starting from approaches based on the Migdal's Finite-Fermi-System theory [1]. In the presented study, the mentioned possibility is realized for each considered parent nucleus by a detailed comparison of the experimental $GT^{(-)}$ strength distribution obtained in a wide excitation-energy interval, that includes GT resonance (GTR), with the respective $GT^{(-)}$ strength function evaluated within the semi-microscopic Particle-Hole Dispersive Optical Model (PHDOM). The model is proved to be useful tool for describing main properties of various Giant

Resonances (GRs) in medium-heavy-mass spherical nuclei (see, e.g., Refs. [2–4] and references therein). In particular, a rather full PHDOM-based description of properties of GTR in the $^{48}$Ca, $^{90}$Zr, $^{132}$Sn, and $^{208}$Pb parent nuclei has been proposed in Ref. [2].

In this brief work, we extend the mentioned study of Ref. [2] by essentially specifying the choice of model parameters. A mean field, the Landau-Migdal p-h interaction, and the imaginary part of the properly parameterized energy-averaged p-h self-energy term responsible for the spreading effect are the phenomenological input quantities employed in implementations of PHDOM. The real part of the mentioned self-energy term is determined by the imaginary part via a proper dispersive relationship. The mean-field parameters are taken from independent (of PHDOM-based studies) data, while the Landau-Migdal parameters (excluding those found from the symmetry conditions) and the "spreading" parameters, which determine the energy dependence of the strength of the imaginary part of the p-h self-energy term (the intensity parameter α, the saturation parameter B, and the gap parameter Δ), are considered as adjustable parameters deduced from a comparison of experimental and calculated GR strength distributions. In the study of Ref. [2], apart from taking the quenching factor Q = 1, the Landau-Migdal parameter *g'* and the "spreading" parameters α and B were adjusted to describe within the model the GTR peak-energy and total width both deduced from the Lorentz-type parameterization of the maximum of the experimental $GT^{(-)}$ strength distribution. The parameter Δ was taken as a universal quantity. In the present study, apart from consideration of this strength distribution in a wide excitation-energy interval, we, also, extend the number of adjustable parameters by inclusion in this number the quenching factor Q. The $GT^{(\mp)}$ strength functions, evaluated within PHDOM by the methods described in detail in Ref. [2], are simply multiplied by the factor Q. For each considered parent nucleus, the four above-mentioned adjustable parameters are deduced from the comparison (performed with the use of the $\chi^2$–method for a wide excitation-energy interval) of the experimental $GT^{(-)}$ strength distribution with respective $GT^{(-)}$ strength function evaluated within PHDOM. Such a comparison is illustrated in Figs. 1 – 4, while the results are presented in Tables I and II. Experimental and calculated $GT^{(-)}$ strength functions for the $^{48}$Ca, $^{90}$Zr, $^{132}$Sn and $^{208}$Pb parent nuclei are given as functions either of ω (the excitation-energy counted off the parent-nucleus ground-state energy), or $E_x = \omega - Q^{(\mp)}$ (the excitation energy of the respective product nucleus with $Q^{(\mp)}$ being the experimental difference of ground-state energies of related product and parent nuclei). The above-mentioned "spreading" parameters determine the energy dependence of the imaginary part of the energy-averaged p-h self-energy term as a three-parametric function of $E_x$ [2].

In Table I, we show the deduced values of Q and g'. Here, we note that: 1) the normalization factor, $C_0$, which determines the Landau-Migdal parameter G' = $C_0$g', is taken as $C_0$ = 300 MeV fm$^3$ [1, 2]; 2) the g' values deduced in Refs. [5 – 8] are shown in Table I after multiplying by factor 4/3 because of using in these references the normalization factor ≈400 MeV fm$^3$ [9]. In Table I, the deduced g' values are compared with: 1) the values deduced in Ref. [2] from the PHDOM-based consideration of GTR maximum only (given in brackets); 2) the values deduced from the continuum-RPA-based consideration of GTR maximum (this consideration is following from PHDOM, neglecting the spreading effect); 3) the values deduced in other studies.

Table I. The deduced Q and g' values (explanations are given in the text).

| Parent nuclei | Q | g' | | |
|---|---|---|---|---|
| | | **PHDOM** | **cRPA** | **Other studies** |
| $^{48}$Ca | 0.60 | 0.75 (0.85) | 0.84 | 0.65 [5] |
| $^{90}$Zr | 0.68 | 0.75 (0.68) | 0.58 | 0.8 ± 0.13 [6]; 0.793 [5] |
| $^{132}$Sn | 0.66 | 0.70 (0.71) | 0.77 | 0.91 ± 0.09 [7] |
| $^{208}$Pb | 0.73 | 0.61 (0.73) | 0.78 | 0.85 [8]; 0.963 [5] |

In Table II, we give the considered excitation-energy intervals $\omega_1 - \omega_2$; values of adjusted "spreading" parameters (the parameters employed in Ref. [2] are given in brackets); values of relative integrated GT strengths (fractions of the Ikeda sum rule), $\mathbf{x}_{12}^{(-)}$ and $\mathbf{x}^{(-)*}$, $\mathbf{x}^* = \mathbf{x}^{(-)*} - \mathbf{x}^{(+)*}$ taken for excitation-energy intervals $\omega_1 - \omega_2$ and 0 – 80 MeV, respectively.

Table II. The considered excitation-energy intervals, adjusted "spreading" parameters, and related fractions parameters (notations are given in the text). The gap parameter is taken as the universal quantity, Δ=3 MeV.

| **Parent Nuclei** | $\boldsymbol{\omega_1 - \omega_2}$, **MeV** | $\boldsymbol{\alpha}$, $\mathbf{MeV^{-1}}$ | **B, MeV** | $\mathbf{x}_{12}^{(-)}$, **%** | $\mathbf{x}^{(-)*}$, **%** | $\mathbf{x}^*$, **%** |
|---|---|---|---|---|---|---|
| $^{48}$Ca | 0.55 – 31.95 | 2.54 (0.25) | 2.21 (5.34) | 86.60 | 92.09 | 84.28 |
| $^{90}$Zr | 6.9 – 56.90 | 0.74 (0.51) | 4.21 (5.24) | 97.72 | 102.02 | 97.56 |
| $^{132}$Sn | 0.0 – 23.30 | 0.27 (0.26) | 6.81 (5.84) | 87.82 | 100.86 | 97.91 |
| $^{208}$Pb | 3.8 – 50.0 | 0.24 (0.24) | 6.05 (5.12) | 100.56 | 102.55 | 100.40 |

We start the comments to obtained results with Figs. 1 – 4. Presented in these figures the PHDOM-based description of experimental GT$^{(-)}$ strength distributions is proved to be satisfactory except for a GTR low-energy distant tail. At least partially, the distinction is explained by ignoring within PHDOM the existence of real low-energy 2p-2h configurations. Within the model, 2p-2h configurations are considered only as doorway-states for the spreading effect. This consideration is valid at sufficiently high excitation energies, when the experimental level density can be described in terms of a statistical model. As for the deduced Q values (Table I), they are in an agreement with the value Q = 0.72 deduced from the description of the experimental nuclear magnetic moments within the Migdal's Finite-Fermi-System Theory [10]. Comments regarding the deduced g' values (Table I) are the following: 1) taking into consideration a large excitation-energy interval in the vicinity of the GTR maximum gives a marked contribution to the deduced value; 2) the microscopically-based phenomenological description of the spreading effect within PHDOM also leads to a marked ("dispersive") contribution to the deduced value; 3) as a rule, a proximity of the g' values, obtained with the use the cRPA limit of PHDOM and the cRPA-based approach employed $\pi + \rho + g'$ model interaction in studies of [5 - 8], is an evidence for a rather small contribution of the one-meson exchange to formation of GTR in parent nuclei under consideration. In this connection, we note the value g' = 1.1 used in Ref. [10] in applying to many medium-heavy-mass spherical nuclei. Turning to the Table II, we first note the large excitation-energy intervals considered, and the reasonable values of calculated fraction parameters. The most interesting points in the Table II are the two sets of adjusted "spreading" parameters. The parameters (given in brackets) deduced with the use of narrow excitation-energy intervals, including the GTR maximum, are only slightly changed with A [2]. The parameters obtained in this work are varied significantly. Being realized in the present study in applying to GTR, the procedure of adjusting the "spreading" parameters with the use of large excitation energy intervals is the first attempt in implementations of PHDOM to describing of various giant resonances. Such attempts are planned to be continued.

In summary, we obtained the values of the Landau-Migdal parameter g' and quenching factor Q from a detailed comparison of experimental Gamow-Teller strength distributions in medium-heavy-mass doubly-closed-shell nuclei and related strength functions evaluated within the semi-microscopic Particle-Hole Dispersive Optical Model. As a rule, the obtained values are close to those obtained in other studies. The presented study is a necessary step for considering within the model the effect of tensor correlations on formation of spin-isospin multipole giant resonances in medium-heavy-mass spherical nuclei. A preliminary study of this problem is given in Ref. [11].

**Acknowledgments.**

The authors are thankful to Profs. M.N. Harakeh and I.N. Borzov for viewing the manuscript and valuable remarks. This work is partially supported by the Program "Priority-2030" for the National Research Nuclear University "MEPhI" and Project No. FSWU-2020-0035 of the Ministry of Science and Higher Education of the Russian Federation (M.H.U.).

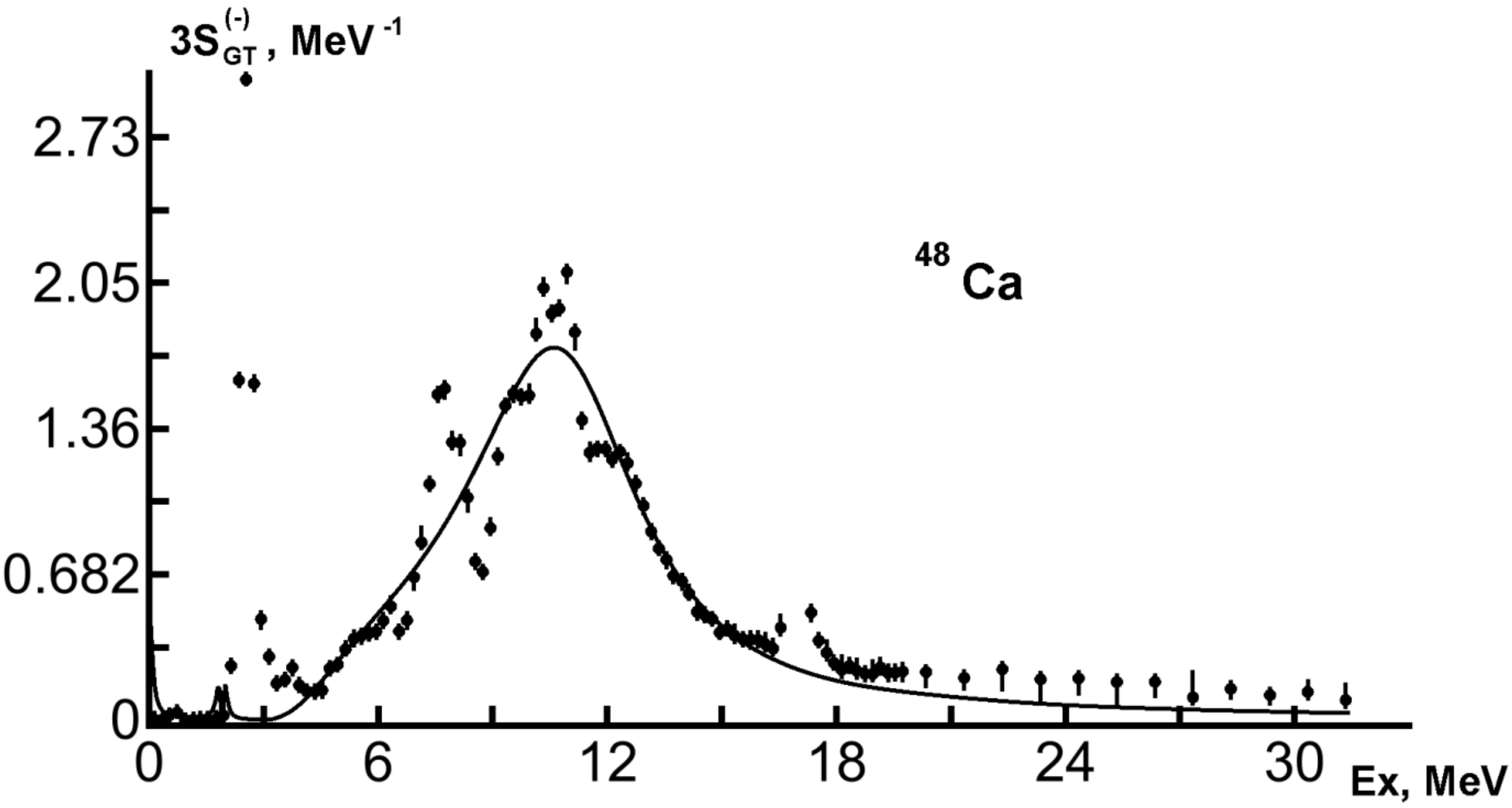

FIG.1. Comparison of the experimental $GT^{(-)}$ strength distribution for the $^{48}$Ca parent nucleus [12] and the respective strength function evaluated within PHDOM with the use of adjusted parameters given in Tables I, II.

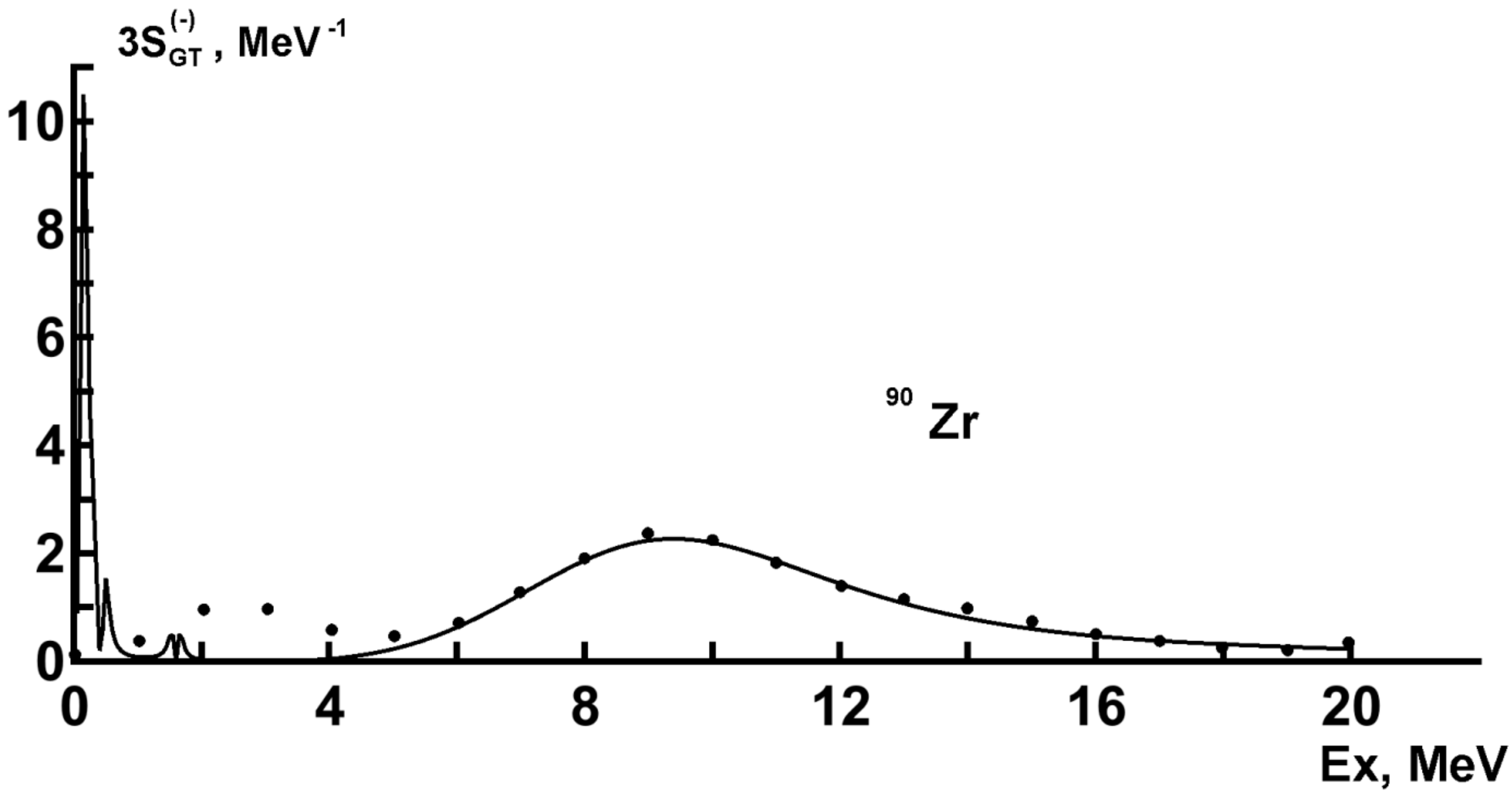

FIG.2. The same as in Fig.1, but for the $^{90}$Zr parent nucleus [13].

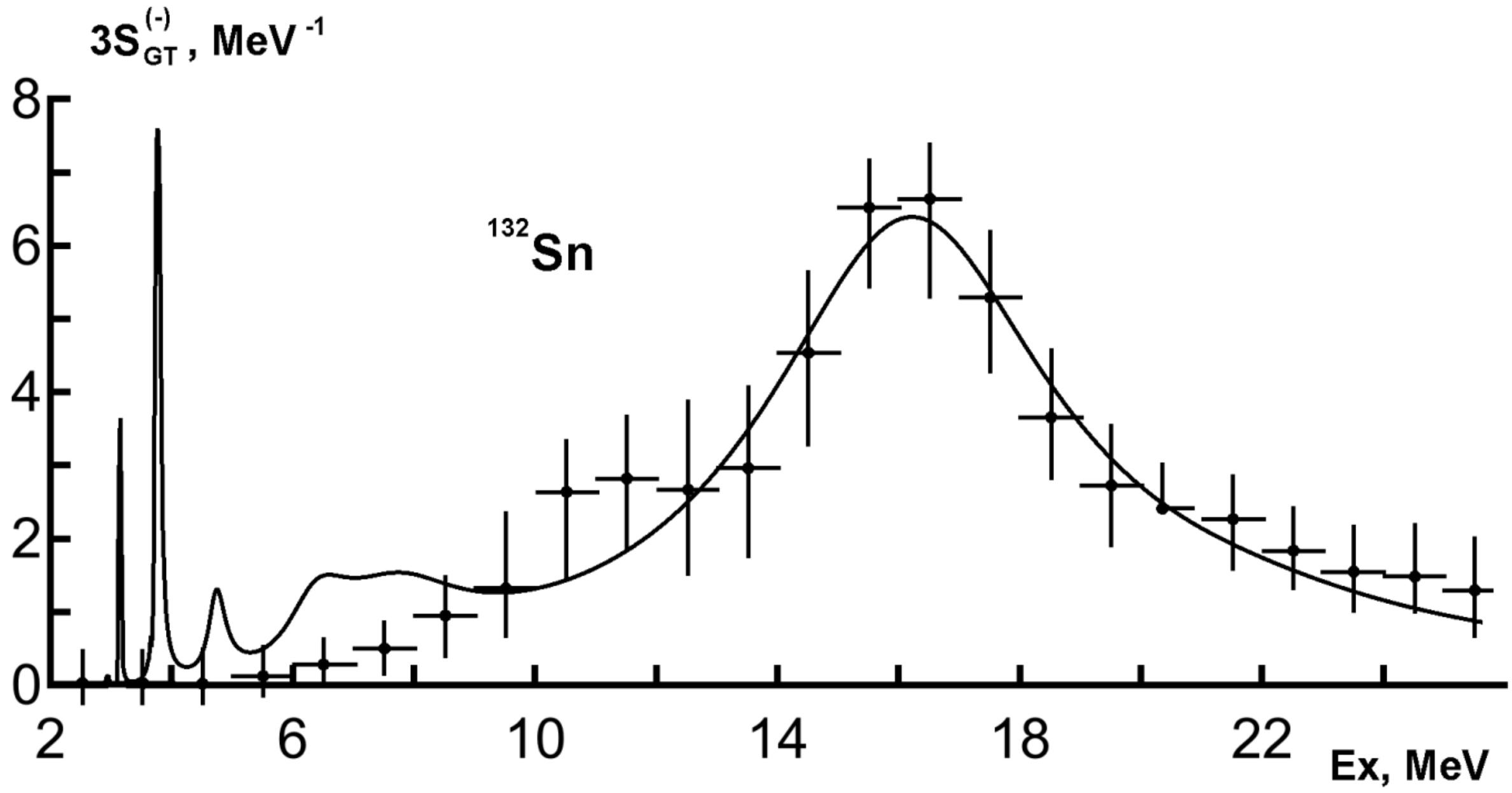

FIG.3. The same as in Fig.1, but for the $^{132}$Sn parent nucleus [7].

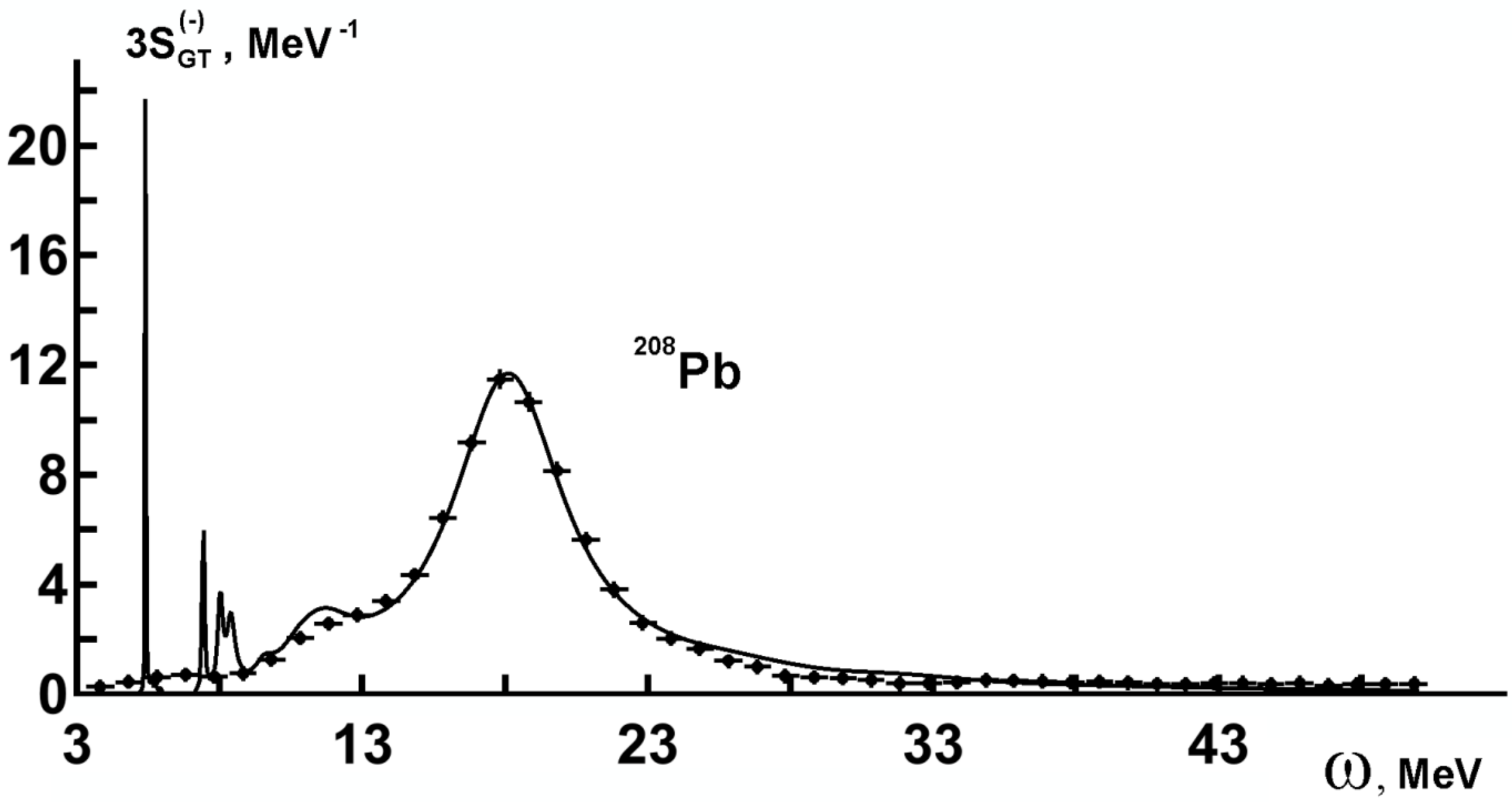

FIG.4. The same as in Fig.1, but for the $^{208}$Pb parent nucleus [8].